\documentclass[english,prl,twocolumn]{revtex4-1}
\usepackage[T1]{fontenc}
\usepackage[latin9]{inputenc}
\usepackage{color}
\usepackage{amsmath}
\usepackage{graphicx}

\makeatletter
\@ifundefined{textcolor}{}
{%
 \definecolor{BLACK}{gray}{0}
 \definecolor{WHITE}{gray}{1}
 \definecolor{RED}{rgb}{1,0,0}
 \definecolor{GREEN}{rgb}{0,1,0}
 \definecolor{BLUE}{rgb}{0,0,1}
 \definecolor{CYAN}{cmyk}{1,0,0,0}
 \definecolor{MAGENTA}{cmyk}{0,1,0,0}
 \definecolor{YELLOW}{cmyk}{0,0,1,0}
 }

\makeatother

\usepackage{babel}

\begin{document}

\title{Unified criteria for multipartite quantum nonlocality}

\author{E. G. Cavalcanti,$^{\text{1}}$ Q. Y. He,$^{\text{2}}$ M. D. Reid,$^{\text{2}}$
and H. M. Wiseman$^{1,\text{3}}$}

\affiliation{$^{\text{1}}$Centre for Quantum Dynamics, Griffith University, Brisbane
QLD 4111\\
$^{\text{2}}$ARC Centre for Quantum-Atom Optics, Swinburne University
of Technology, Melbourne, Australia\\
$^{\text{3}}$ARC Centre for Quantum Computation and Communication
Technology, Griffith University, Brisbane QLD 4111}
\begin{abstract}
Wiseman and co-workers (Phys. Rev. Lett. \textbf{98}, 140402, 2007)
proposed a distinction between the nonlocality classes of Bell nonlocality,
steering and entanglement based on whether or not an overseer \emph{trusts}
each party in a bipartite scenario where they are asked to demonstrate
entanglement. Here we extend that concept to the multipartite case
and derive inequalities that progressively test for those classes
of nonlocality, with different thresholds for each level. This framework
includes the three classes of nonlocality above in special cases and
introduces a family of others.
\end{abstract}
\maketitle
The Einstein-Podolsky-Rosen (EPR) paradox \cite{einstein} revealed
an incompatibility between local causality and the completeness of
quantum mechanics. Bell later discovered that quantum mechanics (QM)
can violate certain inequalities that are predicted by all Local Hidden
Variable (LHV) theories \cite{Bell}. Schrödinger \cite{sch} introduced
the term \emph{steering} to describe the {}``spooky action at a distance''
envisaged by EPR where an agent can apparently ``steer'' a distant
quantum state, and the term \emph{entanglement} for the nonfactorisable
property possessed by all states that demonstrate this phenomenon.

Entanglement, EPR paradox, and Bell's nonlocality were seen as requiring
the same resources until Werner \cite{werner} discovered that not
all entangled states can violate a Bell inequality. At a similar time,
experiments by Ou \emph{et al.} \cite{eprr-1} demonstrated the EPR
paradox, for measurements which would admit a LHV model but based
on criteria specific to the EPR paradox \cite{eprr-2}, thus further
suggesting that the three types of nonlocality embody different physics. 

This idea wasn't fully formalised until recently, when Wiseman, Jones,
and Doherty (WJD) \cite{hw-1} presented a definition of steering
as violation of what they termed a \emph{Local Hidden State }(LHS)
model. They showed that the states exhibiting steering are a strict
\emph{subset} of the entangled states, and a strict \emph{superset}
of the Bell-nonlocal states (those violating a LHV model). In Ref.~\cite{cavalsteer},
the EPR paradox and steering were shown to be equivalent concepts,
and EPR-steering criteria were defined as any experimental consequences,
usually in the form of inequalities, of the LHS model. Based on that
work, Saunders et al. \cite{steerexp} experimentally demonstrated
EPR steering for a Bell-local state, thus confirming EPR steering
as a distinct class of nonlocality. 

The LHS approach provides a rigorous framework from which to derive
a multitude of criteria for the {}``intermediate'' EPR-steering
nonlocality---in scenarios \emph{different} to the one originally
considered by EPR and Schrödinger. This allows fresh investigations
of a form of nonlocality which has a profound historical significance
but about which relatively little is known \cite{opweh-1}. 

However, an important scenario remains unexplored. This prior work
was limited to the bipartite case, as in EPR's original argument.\ WJD
presented an information-theoretic interpretation for the distinction
between entanglement, EPR steering and Bell nonlocality, based on
whether an overseer \emph{trusts} each of two parties in a task where
they are asked to demonstrate entanglement. That approach motivates
our extension to multiple parties.

Further motivation exists, as the $N$-party scenario displays very
interesting behaviour. Greenberger, Horne and Zeilinger (GHZ) \cite{GHZ-1-1}
showed that multipartite {}``GHZ states'' exhibit extreme forms
of nonlocality, which Mermin, followed by Ardehali, Belinski and Klyshko
(MABK) \cite{mermin}, characterised by inequalities whose violation
by QM increases exponentially as $2^{(N-1)/2}$ with $N$. Werner
and Wolf \cite{wernerwolfmaxghz-2} later proved the MABK inequalities
to be those violated by {}``the widest margin in quantum theory'',
for the two-setting, two-outcome experiments. Surprisingly, this multi-party
Bell nonlocality is stable against loss \cite{braunmann-1}. Cabello
et al. \cite{cabello} showed that above a critical detection efficiency
$\eta_{crit}=N/(2N-2)$ there is no LHV model to describe violations
of Mermin's inequality.

In this paper, we introduce a family of locally causal models involving
$N$ parties, $T$ of which are trusted, and derive inequalities to
test for their failure. Entanglement, EPR steering and Bell nonlocality
apply to special cases where all, one or no parties are trusted. These
inequalities take the form of MABK but with different thresholds for
each level of nonlocality. 

We prove that violation of the EPR steering inequalities grows exponentially,
as $2^{(N-2)/2}$, for continuous variable (CV) measurements, and
as $2^{(N-1)/2}$, for dichotomic measurements. Unlike the Bell violation,
however, the dichotomic maximal violation manifests, for all\emph{
}$N$, for \emph{both} of two famous choices of measurement settings:
EPR and GHZ's that gives a perfect correlation between outcomes, and
Bell's, that gives statistical correlation only. Furthermore, we show
as a consequence that EPR steering is more resistant to loss than
Bell nonlocality. 

We begin by introducing some notation: we consider $N$ spatially
separated parties labelled by $j$ who can choose between a number
of experiments. We label the experiments by lower-case letters $x_{j}\in\mathcal{M}_{j}$.
The respective outcomes are labelled by upper-case letters $X_{j}\in\mathcal{O}_{x_{j}}$.
A sufficient specification of the features of the preparation procedure
which are explicitly \emph{known} by the experimenters is labelled
by $\kappa$; a sufficient specification of \emph{any} (possibly unknowable
and thus ``hidden'') variables which may be relevant to the experiments
considered is labelled by $\lambda$. Whenever an equation involving
those variables appears, it is implicitly assumed that the equation
holds for all values of the variables.

Before continuing, we summarise the nonlocality hierarchy for bipartite
systems \cite{hw-1}. The strongest form of nonlocality is \emph{Bell
nonlocality}, in which LHV models are falsified \cite{Bell}. LHV
models require that probabilities for joint measurements at sites
$A$ (Alice) and $B$ (Bob) can be written in the factorisable form:\vspace{-0.2cm}
\begin{multline}
P(X_{A},X_{B}|x_{A},x_{B},\kappa)=\\
\int d\lambda P(\lambda|\kappa)P(X_{A}|x_{A},\lambda,\kappa)P(X_{B}|x_{B},\lambda,\kappa).\label{eq:LHVmodelbipartite}\end{multline}
A\emph{ further} assumption is to require that Bob's {}``local state''
be \emph{quantum}: ie. there must be a quantum state $\rho_{\lambda,\kappa}$
such that for all outcomes $X_{B}$ of all measurements $x_{B}$\begin{equation}
P(X_{B}|x_{B},\lambda,\kappa)=\mathrm{Tr}[E_{X_{B}}\rho_{\lambda,\kappa}]\equiv P_{Q}(X_{B}|x_{B},\lambda,\kappa),\end{equation}
where $E_{X_{B}}$ is the POVM element associated with $X_{B}$. With
this assumption, we arrive at the asymmetric\emph{ }LHS model of WJD
\cite{hw-1}\emph{. EPR Steering} (of Bob's state by Alice) arises
when this model fails \cite{hw-1,cavalsteer}. \emph{Entanglement}
arises as a failure of the Quantum Separable (QS) model, $P(X_{A},X_{B})=\int d\lambda P(\lambda)P_{Q}(X_{A}|\lambda)P_{Q}(X_{B}|\lambda),$
in which one assumes quantum states for \emph{both} systems.

Now consider the following task \cite{hw-1}: Charlie wants to demonstrate
entanglement between $N$ parties. Initially, he will be satisfied
if their correlations cannot be written as a quantum separable (QS)
model (leaving $\kappa,x_{A},x_{B}$ henceforth implicit)\vspace{-0.2cm}
 \begin{equation}
P(X_{1},...,X_{N})=\int d\lambda P(\lambda)\prod_{j=1}^{N}P_{Q}(X_{j}|\lambda).\label{eq:QS_model_multipartite-1-1}\end{equation}
Suppose now that Charlie trusts the first $T$ agents (and their apparata),
but not the remaining $N-T$ (or their apparata). That is, he does
not trust that the measurement outcomes reported by the untrusted
group correspond to the quantum observables they report to have measured.
In this case, the QS model may be violated even without any entanglement.
However, if the observed correlations cannot be reproduced by a model
of form\vspace{-0.2cm}
 \begin{multline}
P(X_{1},...,X_{N})=\\
\int d\lambda P(\lambda)\prod_{j=1}^{T}P_{Q}(X_{j}|\lambda)\prod_{j=T+1}^{N}P(X_{j}|\lambda),\label{eq:LHS_model_multipartite}\end{multline}
with the outcomes of the untrusted parties given by arbitrary (not
necessarily quantum) LHV distributions $P(X_{j}|\lambda)$, Charlie
will be convinced they share entanglement, since then no locally causal
model exists that could be used by an untrusted party to generate
the statistics. We denote the multipartite LHS model (\ref{eq:LHS_model_multipartite})
with $T$ trusted sites and $N-T$ untrusted sites by LHS$(T,N)$.
Violation of the LHS($T,N$) model confirms entanglement in the presence
of $N-T$ untrusted sites.

Violation of a LHS($N,N$) model is equivalent to a standard entanglement
test, while violation of LHS$(0,N)$ model implies Bell nonlocality.
Following Ref.~\cite{hw-1}, violation of a LHS($1,2$) model is
a demonstration of EPR steering. For $T>1$, failure of LHS($T,N$)
implies entanglement, but not necessarily EPR steering. The most interesting
case is violation of LHS($1,N$). This violation can only occur if
EPR steering exists for some bipartition of the $N$ sites: due to
the violation of a LHS($1,2$) model between the trusted site and
the untrusted sites taken as a group, or to the violation of a LHV
model among the non-trusted sites (which in turn implies EPR steering).
However, in these cases we cannot interpret the situation as the state
in a specific site being steered by the others, so we refer to violation
of LHS($1,N$) as a multipartite demonstration of\emph{ }EPR steering\emph{.}

We now turn to derive inequalities to demonstrate failure of each
member of the family of locally causal models (\ref{eq:LHS_model_multipartite}).
Following \cite{mermin,cvbell2-1-1}, we construct complex functions
$F_{j}^{\pm}=X_{j}\pm iY_{j}$ of measurement outcomes $X_{j},Y_{j}$
at each site $j$. For any LHS($T,N$) model (\ref{eq:LHS_model_multipartite}),
$\langle\prod_{j=1}^{N}F_{j}^{s_{j}}\rangle=\int d\lambda P(\lambda)\prod_{j=1}^{N}\langle F_{j}^{s_{j}}\rangle_{\lambda}$,
where $s_{j}\in\{-,+\}$. Here $\langle F_{j}^{\pm}\rangle_{\lambda}=\langle X_{j}\rangle_{\lambda}\pm i\langle Y_{j}\rangle_{\lambda}$
and $\langle X_{j}\rangle_{\lambda}=\sum_{X_{j}}P(X_{j}|\lambda)\, X_{j}$,
with $P(X_{j}|\lambda)=P_{Q}(X_{j}|\lambda)$ for the trusted parties,
$1\leq j\leq T$. From the variance inequality it then follows that\vspace{-0.2cm}
 \begin{equation}
|\langle\prod_{j=1}^{N}F_{j}^{s_{j}}\rangle|^{2}\leq\int d\lambda P(\lambda)\prod_{j=1}^{N}|\langle F_{j}^{s_{j}}\rangle_{\lambda}|^{2}.\label{eq:variance_ineq}\end{equation}
Since $|\langle F_{j}^{\pm}\rangle_{\lambda}|^{2}=\langle X_{j}\rangle_{\lambda}^{2}+\langle Y_{j}\rangle_{\lambda}^{2}$,
it follows from the non-negativity of variances that for any LHV (untrusted)
state: $|\langle F_{j}^{\pm}\rangle_{\lambda}|^{2}\leq\langle X_{j}^{2}\rangle_{\lambda}+\langle Y_{j}^{2}\rangle_{\lambda}$.
For a local \emph{quantum} (trusted) state, quantum uncertainty relations
impose further restrictions. We consider uncertainty relations of
the form $\Delta^{2}X_{j}+\Delta^{2}Y_{j}\geq C_{j}$, where $C_{j}$
depends on the operators associated to $x_{j}$ and $y_{j}$. Substituting
on \eqref{eq:variance_ineq} we obtain a family of nonlocality criteria:\vspace{-0.2cm}
\begin{equation}
|\langle\prod_{j=1}^{N}F_{j}^{s_{j}}\rangle|\leq\left\langle \prod_{j=1}^{T}(X_{j}^{2}+Y_{j}^{2}-C_{j})\prod_{j=T+1}^{N}(X_{j}^{2}+Y_{j}^{2})\right\rangle ^{1/2}.\label{eq:main_ineq}\end{equation}

So far no assumption was made about the measurements $x_{j},y_{j}$.
We now consider two cases: continuous and dichotomic outcomes. For
the continuous case, we assume position-momentum conjugation relations
$[\hat{x_{j}},\hat{y_{j}}]=i$ for the trusted sites, which imply
the local uncertainty relation $\Delta^{2}X_{j}+\Delta^{2}Y_{j}\geq1$,
i.e., $C_{j}=1$. Thus the LHS model (\ref{eq:LHS_model_multipartite})
implies\vspace{-0.2cm}
 \begin{equation}
|\langle\prod_{j=1}^{N}F_{j}^{s_{j}}\rangle|\leq\left\langle \prod_{j=1}^{T}(X_{j}^{2}+Y_{j}^{2}-1)\prod_{j=T+1}^{N}(X_{j}^{2}+Y_{j}^{2})\right\rangle ^{1/2}.\label{eq:multipartiteproofcv}\end{equation}
 With $T=N$, we obtain the entanglement criterion of Hillery and
Zubairy \cite{hillzub-1}; with $T=0$, we obtain the Bell inequality
of Cavalcanti \emph{et al.} \cite{cvbell2-1-1}. We now show these
inequalities can be violated by QM. Using quadrature operators $\hat{x}_{j}=(\hat{a}_{j}+\hat{a}_{j}^{\dagger})/\sqrt{2}$,
$\hat{y}_{j}=i(\hat{a}_{j}^{\dagger}-\hat{a}_{j})/\sqrt{2}$, where
$\hat{a}_{j}^{\dagger},\hat{a}_{j}$ are bosonic creation/annihilation
operators ($[\hat{a}_{j},\hat{a}_{k}^{\dagger}]=\delta_{j,k}$), we
obtain $F_{j}^{+}=\sqrt{2}\hat{a}_{j}^{\dagger}$ and $F_{j}^{-}=\sqrt{2}\hat{a}_{j}$,
and $(\hat{x}_{j}^{2}+\hat{y}_{j}^{2})=2\hat{a}_{j}^{\dagger}\hat{a}_{j}+1=2\hat{n}_{j}+1$,
$\hat{n}_{j}$ being the number operator for each site. Symbolising
$\hat{a}^{+}=\hat{a}^{\dagger}$ and $\hat{a}^{-}=\hat{a}$, the inequalities
(\ref{eq:multipartiteproofcv}) will be violated when\vspace{-0.2cm}
 \begin{equation}
|\langle\hat{a}_{1}^{s_{1}}...\hat{a}_{N}^{s_{N}}\rangle|>\langle\prod_{j=1}^{T}\hat{n}_{j}\prod_{j=T+1}^{N}(\hat{n}_{j}+1/2)\rangle^{1/2}.\label{eq:ataineq}\end{equation}
Consider the $N$ sites prepared in a GHZ-type state \cite{GHZ-1-1}\vspace{-0.2cm}
 \begin{equation}
|\psi\rangle=\frac{1}{\sqrt{2}}(|0\rangle^{\otimes r}|1\rangle^{\otimes N-r}+e^{i\phi}|1\rangle^{\otimes r}|0\rangle^{\otimes N-r}),\label{eq:ghz}\end{equation}
where $r\in\{1,...,N\}$; $|n\rangle^{\otimes r}\equiv\bigotimes_{j=1}^{r}|n\rangle_{j}$;
$|n\rangle^{\otimes N-r}\equiv\bigotimes_{j=r+1}^{N}|n\rangle_{j}$
and $|n\rangle_{j}$ are the eigenstates of $\hat{n}_{j}$. Taking
$\phi=0$, the left-side of \eqref{eq:ataineq} is nonzero when $s_{j}=+$
for all $j\leq r$ and $s_{j}=-$ for all $j>r$; or vice-versa. For
those parameters, $|\langle\prod_{k=1}^{N}\hat{a}_{k}^{s_{k}}\rangle|=1/2$.
For the right-side, note that the ordering of the trusted sites does
not need to coincide with the ordering of the sites on Eq.~\eqref{eq:ghz}.
Inspecting Eq.~\eqref{eq:ataineq} we see that the trusted sites
will annihilate the terms of \eqref{eq:ghz} when their corresponding
state is $|0\rangle$. Thus for all $T\geq2$, $r\neq0$, we can choose
the ordering on state \eqref{eq:ghz} such that both terms will give
zero contribution and inequality \eqref{eq:multipartiteproofcv} will
be violated by the same amount in all cases. For $T=1$, $r\neq0$,
the right-side is $(3^{r-1}/2^{N})^{1/2}$ and we can violate the
inequality for $N\ge3$ . For the optimal case of $r=1$, the ratio
of left to right side becomes $2^{N-2/2}$, an exponential increase
of EPR-steering with $N$. Bell nonlocality ($T=0$) requires $N>9$,
with $r=N/2$ optimal, as shown in \cite{cvbell2-1-1,he cfrd pra-1}.

We now examine the dichotomic case using qubits. We choose $\hat{x}_{j}=\sigma_{j}^{\theta}$,
$\hat{y}_{j}=\sigma_{j}^{\theta+\pi/2}$ where $\sigma_{j}^{\theta}=\sigma_{j}^{x}\cos\theta+\sigma_{j}^{y}\sin\theta$,
and $\sigma_{j}^{x/y}$ are the Pauli spin operators ($\theta$ can
be different for each site). Using the local uncertainty relation
$\Delta^{2}\sigma_{j}^{x}+\Delta^{2}\sigma_{j}^{y}\geq1$ \cite{hoftoth-1}
($C_{j}=1$) for the trusted sites and the identity $(\sigma_{j}^{\theta})^{2}=1$,
inequality \eqref{eq:main_ineq} becomes\vspace{-0.15cm}
\begin{equation}
|\langle\prod_{j=1}^{N}F_{j}^{s_{j}}\rangle|\leq2^{(N-T)/2}.\label{eq:qubit_ineq}\end{equation}
 Defining the Hermitian parts of the operator product by $\prod_{j=1}^{N}F_{j}^{s_{j}}=\mathrm{Re}\Pi_{N}+i\mathrm{Im}\Pi_{N}$,
inequality \eqref{eq:qubit_ineq} implies \begin{eqnarray}
\langle\mathrm{Re}\Pi_{N}\rangle,\langle\mathrm{Im}\Pi_{N}\rangle & \leq & 2^{(N-T)/2},\label{eq:merminsteer}\\
\langle\mathrm{Re}\Pi_{N}\rangle+\langle\mathrm{Im}\Pi_{N}\rangle & \leq & 2^{(N-T+1)/2},\label{eq:merminsteerstat-2}\end{eqnarray}

For $T=N$ these reduce to the separability inequalities of Roy \cite{royprl}.
These inequalities take the form of the MABK inequalities, but for
the Bell case ($T=0),$ a stronger bound can be found \cite{mermin}:
for \emph{odd }$N$ only, \begin{equation}
\langle\mathrm{Re}\Pi_{N}\rangle,\mathrm{\langle Im}\Pi_{N}\rangle\leq2^{(N-1)/2},\label{eq:MABKodd}\end{equation}
 (which is Mermin's inequality), and for \emph{even }$N$ only, \begin{equation}
\langle\mathrm{Re}\Pi_{N}\rangle+\langle\mathrm{Im}\Pi_{N}\rangle\leq2^{N/2}.\label{eq:mabkeven}\end{equation}
(which is the Ardehali-CHSH inequality \cite{Bell,mermin}). The reason
for the different bounds is that in the Bell case, the set of values
of $\langle\prod_{j}F_{j}^{s_{j}}\rangle$ that is defined by all
convex combinations of the classical extreme points (i.e., all LHV
models) is a square in the complex plane \cite{wernerwolfmaxghz-2},
and thus the edges of the square provide tighter bounds than that
given by the maximum modulus of $\langle\prod_{j}F_{j}^{s_{j}}\rangle$.
When one or more parties are treated as having local quantum states,
however, this set becomes a circle, due to the continuum of pure states
allowed by quantum mechanics, and thus the bound given by \eqref{eq:qubit_ineq}
is tight.

Defining $|0\rangle$ and $|1\rangle$ now as eigenstates of $\sigma^{z}$,
the GHZ state \eqref{eq:ghz} ($r=N)$ violates (\ref{eq:merminsteer}-\ref{eq:mabkeven})
by the maximum amount for QM \cite{wernerwolfmaxghz-2}. This occurs
for Mermin-type inequalities (\ref{eq:merminsteer}), (\ref{eq:MABKodd})
for $F_{j}=\sigma_{j}^{x}\pm i\sigma_{j}^{y}$ ($j=1,...,N$), which
is the case of the EPR-Bohm and GHZ paradoxes \cite{einstein,GHZ-1-1}
that yield perfect correlations between spatially separated spins.
The quantum prediction is:\begin{equation}
\langle\mathrm{Re}\Pi_{N}\rangle,\mathrm{\langle Im}\Pi_{N}\rangle=2^{N-1}.\label{eq:qm1}\end{equation}
With $F_{j}=\sigma_{j}^{x}-i\sigma_{j}^{y}$ ($j\neq N$), $F_{N}=\sigma^{-\pi/4}+i\sigma^{\pi/4}$
we get the maximum quantum prediction for the Ardehali-type inequalities
(\ref{eq:merminsteerstat-2}), (\ref{eq:mabkeven}):\begin{equation}
\langle\mathrm{Re}\Pi_{N}\rangle+\langle\mathrm{Im}\Pi_{N}\rangle=2^{N-1/2}.\label{eq:qm2}\end{equation}
The ratio of left to right side of (\ref{eq:merminsteer}) and (\ref{eq:merminsteerstat-2})
is thus $S_{N}=2^{(N+T-2)/2}$, an exponential growth for all $T$. 

EPR steering is shown when the inequalities (\ref{eq:merminsteer})
and (\ref{eq:merminsteerstat-2}) with $T=1$ are violated. Interestingly,
while the ratio $S_{N}=2^{(N-1)/2}$ is unchanged from the optimal
MABK case, there is the new feature that this ratio is achieved for
\emph{both} statistical and perfect correlations, i.e. via \emph{both}
inequalities (\ref{eq:merminsteer}) and (\ref{eq:merminsteerstat-2}).
Thus, additional EPR-steering criteria different to the MABK inequalities
follow from (\ref{eq:merminsteerstat-2}) when $N$ is odd, and from
(\ref{eq:merminsteer}) when $N$ is even: e.g. EPR steering is confirmed
if $|\langle\sigma_{1}^{x}\sigma_{2}^{x}\rangle-\langle\sigma_{1}^{y}\sigma_{2}^{y}\rangle|>\sqrt{2}$.
As shown by Roy \cite{royprl}, entanglement criteria also follow
from both inequalities when $T=N$: e.g. entanglement is confirmed
for the CHSH Bell inequality \cite{Bell} with a lower theshold: $|\langle\sigma_{1}^{x}\sigma_{2}^{x'}\rangle-\langle\sigma_{1}^{y}\sigma_{2}^{y'}\rangle+\langle\sigma_{1}^{y}\sigma_{2}^{x'}\rangle+\langle\sigma_{1}^{x}\sigma_{2}^{y'}\rangle|>\sqrt{2}$.

\begin{figure}
\textcolor{black}{\includegraphics[width=1\columnwidth]{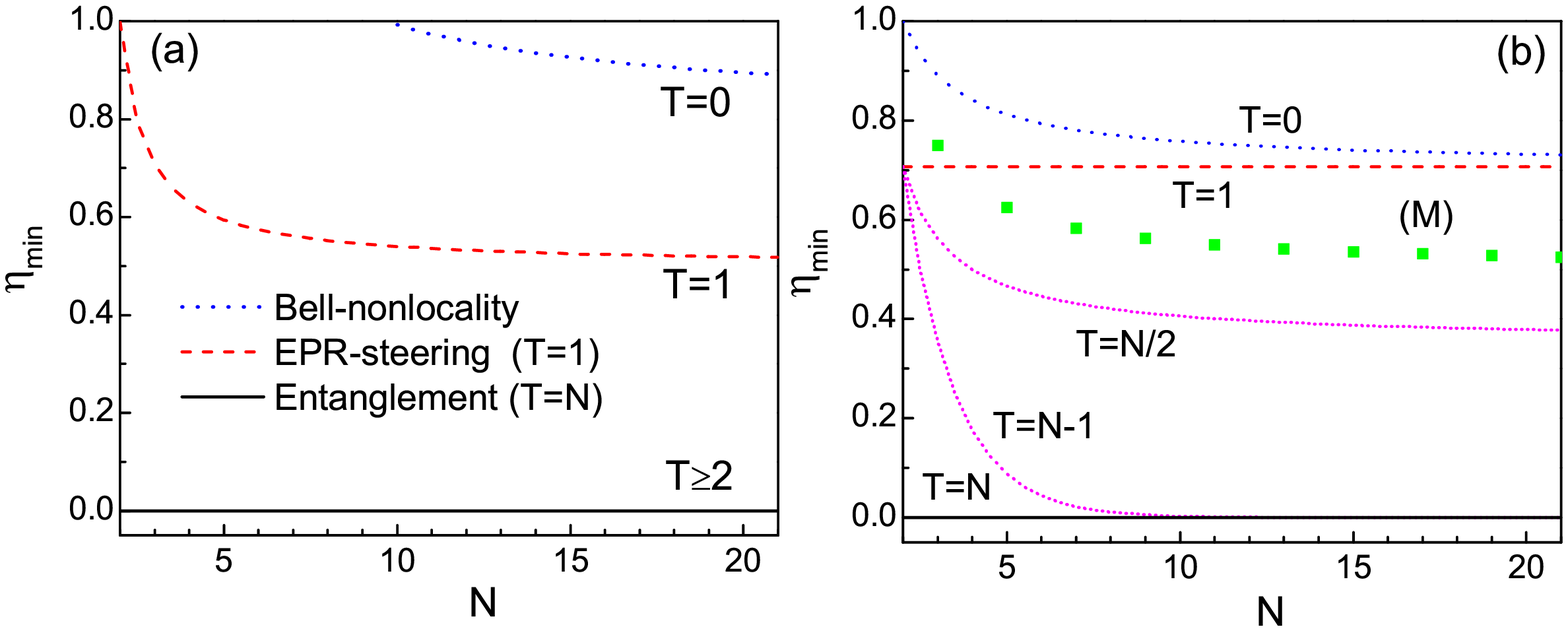}}

\caption{Efficiency $\eta$ sufficient to demonstrate nonlocality using (\ref{eq:multipartiteproofcv})
and (\ref{eq:merminsteer})--(\ref{eq:mabkeven}) with states (\ref{eq:ghz}).
(a) CV case. (b) Dichotomic case. For $T\geq1$, $\eta_{{\rm min}}$
values are for \emph{both} inequalities (\ref{eq:merminsteer}) and
(\ref{eq:merminsteerstat-2}), but for $T=0$ apply \emph{only} to
(\ref{eq:MABKodd}) for odd $N$, and to (\ref{eq:mabkeven}) for
even $N$. Points (M) give $\eta_{{\rm min}}$ necessary and sufficient
for failure of LHV using Mermin's inequality (\ref{eq:MABKodd}).
}

\end{figure}

Crucial in many nonlocality experiments is the effect of loss \cite{cabello}.
Following \cite{ystbeamsplit-1}, we model loss with a beam-splitter
and calculate moments of detected fields, using $a_{det}=\sqrt{\eta}a+\sqrt{1-\eta}a_{vac}$.
Here $a_{vac}$ is the operator for a vacuum reservoir mode into which
quanta are lost.

To summarise the calculation, we start with the continuous variable
case: $[a_{det},a_{det}^{\dagger}]=1$ and inequalities (\ref{eq:multipartiteproofcv}-\ref{eq:ataineq})
still apply. The left-side of (\ref{eq:ataineq}) becomes $\eta_{t}^{T/2}\eta_{u}^{(N-T)/2}/2$
where $\eta_{t}$ ($\eta_{u}$) are efficiencies at trusted (untrusted)
sites respectively. Examining the right-side and optimising $r$ of
\eqref{eq:ghz}, we can detect violation of LHS($T,N$) for $T\geq2$,
with $r\geq2$ and \emph{any} nonzero efficiencies $\eta_{t},\eta_{u}$.
To test EPR steering ($T=1$) we need (with $r\geq1$) $\eta_{u}^{N-1}>2^{r-N+1}(\eta_{u}+1/2)^{r-1}$,
which reduces to\begin{equation}
\eta_{u}>2^{1/(N-1)}/2\end{equation}
for $r=1$. We note an interesting asymmetry: there is sensitivity
to loss at the \emph{untrusted} sites only, an effect noticed for
the bipartite EPR paradox \cite{rrmp}. The limiting efficiency of
50\% required to demonstrate EPR steering is much more accessible
than that required for Bell nonlocality \cite{cvbell2-1-1} ($\eta_{u}>(1+\sqrt{5})/4\approx81\%$
as $N\rightarrow\infty$). 

For the qubit case, violation of inequalities (\ref{eq:merminsteer})--(\ref{eq:mabkeven})
is the same for EPR steering and Bell nonlocality. However, analysis
reveals that in important scenarios the former will be less sensitive
to loss. As GHZ states have been prepared for qubits in optical polarisation
states \cite{ghzexp}, \textcolor{black}{we model loss with beam-splitters
as in the continuous-variable case. Denoting mode operators by $a_{\pm}$,
we introduce Schwinger spins \cite{tothsch} for each site $j$ (subscripts
$j$ for operators are implicit): $s^{z}=a_{+}^{\dagger}a_{+}-a_{-}^{\dagger}a_{-},$
$s^{x}=(a_{+}^{\dagger}a_{-}+a_{+}a_{-}^{\dagger}),$ $s^{y}=(a_{+}^{\dagger}a_{-}-a_{+}a_{-}^{\dagger})/i,$
and $s^{2}=n(n+2),$ where $n=a_{+}^{\dagger}a_{+}+a_{-}^{\dagger}a_{-}$
is the number operator for each site. We map the components of the
GHZ state \eqref{eq:ghz} }into the $\pm1$ eigenstates of $s^{z}$:
$|0\rangle_{j}\rightarrow|0\rangle_{+j}|1\rangle_{-j}$ and $|1\rangle_{j}\rightarrow|1\rangle_{+j}|0\rangle_{-j}$.\textcolor{black}{{}
With loss of a photon at site $j$, the local state becomes }$|0\rangle_{+j}|0\rangle_{-j}$\textcolor{black}{,
and $s^{z}=0$. The Bell inequalities (\ref{eq:MABKodd})- (\ref{eq:mabkeven})
still follow from \eqref{eq:main_ineq} with the extra `0' outcome,
since $X^{2},Y^{2}\leq1$.} For $T>0$, we use the uncertainty relation
$\Delta^{2}s^{x}+\Delta^{2}s^{y}\geq\Delta^{2}n-\Delta^{2}s^{z}+2\langle n\rangle$
\cite{tothsch} to obtain $|\langle F_{j}^{\pm}\rangle_{\lambda}|^{2}\leq\eta_{t}^{2}$
\cite{proof}.\textcolor{black}{{} }Inequality \eqref{eq:main_ineq}
becomes\begin{equation}
{\color{black}|\langle\prod_{j=1}^{N}F_{j}^{s_{j}}\rangle|\leq2^{(N-T)/2}\eta_{t}^{T}.}\label{eq:spin_ineq_loss}\end{equation}
 With state \eqref{eq:ghz} ($r=N$), we need $\eta_{u}>2^{(2-N-T)/(2(N-T))}$
to detect failure of the LHS($T,N$) model via these inequalities.
EPR steering ($T=1$) requires $\eta_{u}>1/\sqrt{2}$ (Fig.1 (b)).
We see that for $N=2,\,3$, EPR steering is detectable with lower
efficiency than  the $\eta_{crit}>N/(2N-2)$  necessary and sufficient
for incompatibility of the measurements of (\ref{eq:merminsteer})
with a LHV model \cite{cabello}. 

In conclusion, we have presented a unified framework to derive criteria
for a family of nonlocality models based on differing levels of trust
on different parties. Those criteria are sufficient to demonstrate
Bell nonlocality, EPR steering and entanglement as special cases.
 The criteria follow from one general proof, with different bounds
for each type of nonocality, thus clarifying the relationship between
them. We note that violation of the inequalities presented here (as
for MABK inequalities) are sufficient but not necessary conditions
for a given quantum state to display the corresponding type of nonlocality.
This is evident on examining the criteria for the mixed Werner states
\cite{werner}, for which the boundary for entanglement and steering
is known \cite{Peres-1,hw-1}. A promising avenue is to search for
criteria involving more than two settings. 

We acknowledge support from the ARC Centre of Excellence program (CE11E0096
and COE348178) and an ARC Postdoctoral Research Fellowship.


\begin{thebibliography}{26}
\bibitem{proof}We rearrange {\footnotesize $\Delta^{2}s_{j}^{x}+\Delta^{2}s_{j}^{y}\geq\Delta^{2}n_{j}-\Delta^{2}s_{j}^{z}+2\langle n_{j}\rangle$}
 and substitute {\footnotesize $s^{2}=n(n+2)$} to get {\footnotesize $\langle s^{x}\rangle^{2}+\langle s^{y}\rangle^{2}\leq\langle n\rangle^{2}-\langle s^{z}\rangle^{2}\leq\langle n\rangle^{2}=\eta_{t}^{2}$}.
\end{thebibliography}
\end{document}